\documentstyle{aipproc}

\begin{document}

\begin{flushright}
\hfill {\rm UR-1622}\\
\hfill {\rm ER/40685/960}\\
\hfill {\rm January 2001}\\
\end{flushright}
\vspace*{2\baselineskip}

\title{Top Quark Physics:  Summary}

\author{Lynne H.~Orr$^*$}
\address{$^*$Department of Physics and Astronomy\\
University of Rochester\\
Rochester, NY 14627-0171}

%\lefthead{LEFT head}
%\righthead{RIGHT head}
\maketitle

\begin{abstract}
This  talk summarizes recent progress in top quark
physics studies for  high energy linear electron-positron colliders 
as presented
at the LCWS2000 Workshop at Fermilab. 
New results  were presented for  top pair production at 
threshold and in the continuum, as well as for top production at
$\gamma\gamma$ colliders.

\end{abstract}

\section*{Introduction}

Top quark production promises to provide an excellent system in which
to study the physics of the Standard Model and beyond, and a linear 
$e^+e^-$ collider
provides an excellent environment in which to do so.  The top quark's 
mass of 174 GeV gives it a number of unique features.   Its decay 
width $\Gamma_t=1.4\ {\rm GeV}$ is so large that top decays
before hadronizing \cite{lifetime}, unlike the lighter quarks.  This rapid 
decay
causes the top quark's spin information  to be passed to its decay 
products. Top also has the largest
coupling to the Higgs sector of all the fermions; indeed the fact that its 
Yukawa coupling is very close to unity suggests that the top quark may play a 
special, fundamental role in Electroweak Symmetry Breaking.   The importance 
of the top quark to new physics studies is indicated by the fact that 
the top quark was featured in talks at this conference not only in the 
present session but also in the Higgs, Electroweak, New Phenomena, and 
Supersymmetry sessions.  This underscores the point that in order to 
get at physics beyond the SM, we must understand the physics of the 
top quark in detail.  

Here we summarize the work that was presented in the Top Quark session.
Significant progress has been made in a number of areas since LCWS99 at 
Sitges (see \cite{sitgesth} and \cite{sitgesexpt} for summaries of 
theoretical and experimental aspects of top physics at LCWS99), including
top production at threshold, at high energies, and in $\gamma\gamma$ collisions.
Threshold studies by the European Working Group were not presented here
but will be included   in the TESLA TDR to be 
released in Spring 2001 \cite{tesla}.

\section*{Top at Threshold}

A linear $e^+e^-$ collider operating in the threshold region for top pair
production (around 350 GeV in the center of mass) allows us to make 
precise measurements of the top quark  mass and the strong coupling 
constant $\alpha_s$.   The top width and couplings can be measured to a 
lesser extent.    Top-antitop pairs cannot form toponium because they 
decay too fast, but there is still a Coulomb-like QCD interaction between the 
$t$ and $\bar t$ which results in notrivial structure in the top production
cross section at 
threshold.  Experimental simulations indicate that a threshold scan could 
give a top mass measurement with an experimental uncertainty
as low as 40 or 50 MeV \cite{sitgesexpt}. 
This would be an improvement over the LHC, where we can expect to measure 
the top mass to a GeV or two at best.
This presents a challenge to
theorists to achieve comparable precision in  threshold cross section
predictions.  

\subsection*{Theoretical Issues}
NNLO calculations of the top threshold cross section were presented at Sitges,
and the results did not at first bode well for the prospect of achieving the desired
theoretical precision \cite{sitgesth}.  
There were several problems:  the NNLO corrections to the normalization were 
unexpectedly large,
the $1S$ peak position (used to determine $m_t$) shifted with respect to 
NLO by large amounts
(600--800 MeV), and the renormalization scale dependence was no better
at NNLO than at NLO.  

The situation was already improving around the time of the Sitges meeting,
and it has been further clarified since then.
 The resolution of the 
peak position--mass problem was
discussed in talks here by Sumino \cite{sumino} and Yakovlev \cite{yakovlev},
and a summary of the situation can be found in \cite{threshsum}. The
problem arose from  the top mass definition used in the calculations.  
Specifically, if one calculates in terms of the familiar pole mass
---  the position of
the pole in the top quark propagator and  the mass measured in 
top quark reconstruction (for example by CDF and D$\emptyset$) ---
one is sensitive to effects associated with long distance/low energy 
phenomena such as hadronization.  This
so-called renormalon ambiguity gives rise to uncertainties of order
$\Lambda_{QCD}$ and results in a shift in the NNLO peak position.

The solution is to switch to a mass definition that is appropriate 
for short distances, such as the $1S$ mass; see \cite{threshsum} for a 
discussion of the various possibilities.  
The renormalon contribution to the relation 
between the pole mass and the short distance mass is exactly canceled by a 
corresponding contribution in the relation between pole mass and 
the binding energy of the $t\bar{t}$ state.  The resulting NNLO peak
shift is well under control, and one can realistically expect much-improved
theoretical precision.

Just what it would take to achieve 50 MeV theoretical precision was the 
subject of the talk by Sumino \cite{sumino}.   It requires higher-order 
calculations in which the renormalons continue to cancel.  There was some
confusion  at Sitges about order counting in the cancellation, but this has since
 been clarified:  the 
cancellation takes place at {\it different} orders in $\alpha_s$ in the 
binding energy and the $m_{pole}$--$m_{\bar{MS}}$ relation; one has to go one
order higher in $E_{binding}$.   For the desired theoretical precision, one needs
${\cal{O}}(\alpha_s^4)$ in the mass and ${\cal{O}}(\alpha_s^5)$ in 
$E_{binding}$.  Sumino reported results for $E_{binding}$ to 
${\cal{O}}(\alpha_s^5)$ in the large-$\beta_0$ approximation, which is sufficient 
to demonstrate the renormalon cancellation.  He obtained
the following masses for the   ``toponium'' $1S$ state (keeping the contributions
from individual orders in $\alpha_s$ separate):
\begin{eqnarray}
M_{1S} &=& 2 \times (174.79  -  0.46  - 0.39  - 0.28  -  0.19)\ {\rm GeV}\\
        &=& 2 \times (165.00  +  7.21 + 1.24  + 0.22  +  0.052)\ {\rm GeV}
\end{eqnarray}
where the first result is for the pole-mass scheme and the second is for the 
$\bar{MS}$-mass scheme; the improvement in convergence in the latter scheme 
is clear.  The last number in each line is a new result reported here at 
LCWS2000.
This represents significant progress; still needed for the desired 
theoretical precision are the ${\cal{O}}(\alpha_s^4 m)$ relation between 
$m_{pole}$ and $m_{\bar{MS}}$ (a four-loop calculation),  inclusion of 
final-state interactions and electroweak corrections, and a determination of 
the effect of initial state radiation on the position of the $1S$ peak.

Yakovlev also discussed top production at the threshold \cite{yakovlev}.
He introduced a new short-distance mass definition, the $\bar{PS}$ 
mass, which is a modification of the $PS$ (potential-subtracted) mass
defined in \cite{psmass}.  The $\bar{PS}$ mass includes recoil corrections of order
$1/m$ and further improves the behavior of the cross section.  Yakovlev
also reported that nonfactorizable corrections to the threshold cross section
cancel at NNLO.

The peak position problems with the top threshold are therefore now solved, paving the 
way for a precise threshold mass measurement. The problems associated
with the overall normalization
and renormalization scale  dependence must also be resolved if we  wish
to measure top couplings at the threshold.  Fortunately, there was some good news on 
this front about the time of  this workshop.  Although it was not
presented here, a new threshold
calculation appeared recently which involves resumming  logarithms of 
the top velocity.  The calculation results in a vast reduction of the
theoretical uncertainty to the few percent level.  Details can be found in
\cite{threshnew}.

\subsection*{Experimental Issues}

Ikematsu discussed experimental aspects of top momentum reconstruction 
near threshold \cite{ikematsu}.  Previous studies have focused on threshold 
scans (see
\cite{sitgesexpt} for a status report) which measure the total top
production cross section.  But as Ikematsu pointed out, 
we can extract additional information by examining threshold top events in more
detail.  For example, anomalous CP-violating gluon EDM couplings are 
enhanced at the threshold.  To study these, we need to measure the momentum
of the top quark.  Ikematsu presented results of an experimental simulation
at $\sqrt{s}=2m_t+2\ {\rm GeV}$ including ISR, beamstrahlung, and QCD
threshold corrections for lepton plus jets events.  He compared a traditional
direct momentum reconstruction algorithm to a kinematic fitting technique that 
incorporated a likelihood function.  The kinematic fit significantly
improves energy resolution and resolution in the direction of the top 
momentum.  Extension of the study to energies below threshold and 
incorporation of an improved jet clustering algorithm are in progress.

\section*{Continuum Top Production}

\subsection*{Anomalous Couplings}

At center of mass energies well above threshold, a linear collider allows 
us to probe the couplings associated with the top production and decay
vertices.  The
production vertex involves the couplings of top to the photon and $Z^0$, 
which are difficult-to-impossible to measure at the LHC.  The  decays
are sensitive to the coupling of top to the $W$ boson and $b$ quark.
In both of these processes we can look for CP-violating couplings in electric
and weak dipole moments.  The work reported here \cite{kiyo,iwasaki}
focused on the top production vertex.

Kiyo \cite{kiyo} discussed the effects of anomalous $\gamma$ and 
$Z^0$  couplings in spin correlations 
in top production and decay.  Using the off-diagonal spin basis of Parke and 
Shadmi \cite{offdiag}, Kiyo showed that the azimuthal angle dependence of 
final-state leptons can be very sensitive to CP-violating effects.  
Let the coupling at the top production vertex be given by 
\begin{equation}
g_{tt\gamma,Z}= g_V\{Q_L^{\gamma,Z}\gamma_\mu L + Q_R^{\gamma,Z}\gamma_\mu R
              +{{(p_t - p_{\bar{t}})_\mu}\over{2m_t}} [G_L^{\gamma,Z}L
                   +G_R^{\gamma,Z}R]\}
\end{equation}
with anomalous couplings $G_{L,R}$.  With $if_3=G_R-G_L$, $|f_3|=1$ corresponds
to an electric dipole moment of $10^{-16} e\; {\rm cm}$.  
One can define a leptonic azimuthal asymmetry as a function of the top quark
scattering angle.  This asymmetry  vanishes in the SM but can be as large
as 5 or 10\% for $|f_3|$ of the order of 0.2.  
In addition, one can distinguish between the various couplings ($\gamma$ vs.\ $Z$
and positive vs.\ negative)
by adjusting the polarization of the initial electrons and positrons.

This method does require reconstruction of the top quark direction.
Being able to distinguish the up- vs. down-type decay products  of the $W$
boson is also helpful;  in leptonic $W$ decays this information is 
given by the lepton charge.

M.~Iwasaki addressed these experimental issues associated with 
measuring the couplings at the top production vertex \cite{iwasaki}.
The particular questions of interest were how well we can reconstruct the
$t$ and $\bar{t}$ quarks (and identify which is which), whether we can
identify $I_z$ of the $W$ decay products, and to what extent we can use 
this information to constrain couplings.  She presented results from an 
LCD fast simulation using Pandora-Pythia, with ISR, beamstrahlung, and QCD 
radiation included.  

The presence of two neutrinos in dilepton events makes it difficult to
obtain the $t$ and $\bar{t}$ momenta, so this mode was not included in 
Iwasaki's studies.
For lepton plus jets events, jet energies were obtained with 
a combination of tracking (for charged particles) and calorimetry
(for neutrals) rather than straight calorimetry.  $b$-tagging was performed 
by requiring that the $p_T$-corrected mass at the vertex be greater than
1.8 GeV, and by cutting on the number of significant tracks in the jet.
The resulting flavor tag dramatically improved $W$ and top mass reconstruction.
The lepton charge provided $I_z$ information.

Flavor tagging played an even more important role in the all-jets mode, 
with six jets and no leptons.  Here Iwasaki used the mass tag for $b$ tagging 
and added a measurement of vertex charge to distinguish $b$ from $\bar{b}$.
A charm tag --- using a combination of $p_T$ corrected mass and vertex
momentum --- was also shown to be reasonably effective (28\% efficiency and
69\% purity) in obtaining detailed information about the $W$ decays.  Note
that such a charm tag would be very difficult at the LHC.

The result of the lepton plus jets analysis is that we can expect measurements 
to about
$2$--$4\times 10^{-2}$ in the axial couplings $F^{\gamma,Z}_{1A}$.  Vector
coupling studies are in progress, as studies of the top decay vertex.
An analysis for the all-jets mode is also on the way.

\subsection*{QCD}

Precision measurements require precision predictions that include QCD 
corrections.  Progress was reported here on fixed-order QCD corrections to top 
production and decay \cite{macesanu} and improvements in parton shower Monte 
Carlos to take the top mass into account \cite{corcella,sjostrand}.

QCD corrections to top processes have previously been performed 
for the production and decay processes separately  or combined in the 
on-shell approximation.  As of the Sitges meeting \cite{sitgesth},
real gluon corrections to  the full top production and decay process with 
off-shell top quarks were calculated.
Macesanu discussed the calculation of virtual 
corrections to top production and decay    for off-shell
top quarks \cite{macesanu} with all interferences, including so-called 
nonfactorizable 
corrections that contain diagrams with gluons connecting different parts
of the process ($t$ production and $\bar{t}$ decay, for example).  The
computation is being performed in the double-pole approximation, keeping 
only contributions from terms which contain two resonant top quarks, in
analogy to the RacoonWW approach \cite{wackeroth} to corrections to
$W$ pair production.  The loop corrections are nearly complete and will be
combined with the result for real gluon radiation \cite{real} to give a 
complete NLO description of off-shell top production and decay.

In Monte Carlo programs,  gluon radiation  is often treated in the 
parton shower approximation, using parton splitting functions.  
This involves soft and collinear approximations for the 
radiated gluons.  The shower approximation
incorporates the leading QCD effects to high orders in
$\alpha_s$ and  works well for massless particles.  
Radiation from massive particles like the top quark is more complicate,
however.
In particular,  from massless particles is characterized by a collinear 
singularity, whereas the 
mass of a heavy particle suppresses collinear radiation, creating a ``dead
cone'' in the vicinity of the radiating particle.   
Parton showers have difficulty approximating the dead cone, and in 
addition they have difficulty populating large angles.  

The solution 
is to incorporate matrix element corrections to the parton shower 
approach, and corrections to the programs HERWIG \cite{corcella} and 
Pythia \cite{sjostrand} were reported
at this workshop.  Corcella  showed results for matrix element corrections 
to top decays in HERWIG;  the program now reproduces the matrix
element calculation for radiation in  top events at linear colliders
not too far above threshold.  Corrections for radiation in top production
are in progress, for hadron colliders as well.  Also
underway are the incorporation of spin correlations, and  ISR
and beamstrahlung for linear colliders.  

Sj\"ostrand reported on improvements to Pythia \cite{sjostrand}, including 
matrix element corrections to radiation from {\it all} heavy particles:  not 
only for top and the other SM particles but for supersymmetric particles 
as well.  For radiation in top production at $E_{cm}=500\ {\rm GeV}$,
the corrections increase emissions off the $t\bar{t}$ pair but decrease 
emissions in the top decays, for a net decrease in the entire production
and decay process.  An interesting point to note about the corrections
is a strong dependence 
of the radiation on the spin structure of the underlying process, in 
constrast to the universal behavior of massless splitting functions.

Corcella also discussed  briefly some studies of top mass reconstruction
in dilepton events using HERWIG and Pythia \cite{corcella}.  The lepton energy and 
$bl$ invariant mass spectra were not very sensitive to $m_t$, but the
$E_b$ spectrum near threshold appeared more promising.  However he also showed 
that differences between HERWIG and Pythia can amount to hundreds
of MeV, which already saturates the desired uncertainty in the mass.

\section*{Top at $\gamma\gamma$ Colliders}

Finally, Boos gave a very nice review of top physics at $\gamma\gamma$ 
colliders \cite{boos}. Many processes and measurements are similar to those at 
$e^+e^-$ colliders.  The pair production cross section can be higher
at $\gamma\gamma$ colliders for some polarizations and center of mass energies,
and NLO corrections increase the $\gamma\gamma$ cross section.   The
threshold has similar NNLO problems to $e^+e^-$ colliders, but these
presumably can also be resolved.

A variety of top couplings and non-SM effects can be studied at a 
photon collider.  
The $\gamma tt$ coupling, for example, enters top pair production to 
the fourth power, and is already separate from the top coupling to the $Z$.
Boos showed that  CP properties of the Higgs boson can be studied, as can 
effects of large extra dimensions.  Single top production probes
the $Wtb$ vertex and is sensitive to some technicolor models; ironically, its biggest
background is $t\bar{t}$ pair production.

\section*{Summary}

We have seen at this workshop a lot of progress in top studies as we move 
from looking primarily at the big picture to performing more detailed analyses.  
Some very good news was that the theory of top production at  threshold is 
once more under control.  In continuum production, calculations of
QCD corrections are becoming more sophisticated, and anomalous couplings 
are being studied in more detail.  In addition, the experimental side
of top reconstruction is becoming better understood.  Much work is still under
way, so we anticipate even more progress soon.


\begin{references}


\bibitem{lifetime} I.I.~Bigi, {\it et al.}, 
Y.L.~Dokshitzer, V.~Khoze, J.~Kuhn and P.~Zerwas,
Phys.\ Lett.\  {\bf B181}, 157 (1986); L.H.~Orr and J.L.~Rosner,
Phys.\ Lett.\  {\bf B246}, 221 (1990), {\bf 248} (1990)
474(E).
\bibitem{sitgesth}
A.~H.~Hoang, proceedings of the 4th International Workshop Detectors on Linear 
Colliders (LCWS 99), Sitges, Barcelona, Spain, 28 Apr.\ -- 5 May 1999
hep-ph/9909414.
\bibitem{sitgesexpt}
M.~Martinez, proceedings of the 4th International Workshop Detectors on Linear 
Colliders (LCWS 99), Sitges, Barcelona, Spain, 28 Apr.\ -- 5 May 1999.
\bibitem{tesla} See, for example 
\verb+http://www.desy.de/conferences/ecfa-desy-lc98.html+ for more 
information.
\bibitem{sumino}
Y.~Sumino, these proceedings.
\bibitem{yakovlev}
O.~Yakovlev, these proceedings.
\bibitem{threshsum}
A.~H.~Hoang {\it et al.}, 
Eur.\ Phys.\ J.\ direct{\bf C3}, 1 (2000).
\bibitem{psmass}
M.~Beneke,
Phys.\ Lett.\ {\bf B434}, 115 (1998).
\bibitem{threshnew}
A.~H.~Hoang, A.~V.~Manohar, I.~W.~Stewart and T.~Teubner,
hep-ph/0011254.
\bibitem{ikematsu}
K.~Ikematsu, these proceedings.
\bibitem{kiyo}
Y.~Kiyo, these proceedings.
\bibitem{offdiag}
S.~Parke and Y.~Shadmi,
Phys.\ Lett.\ {\bf B387}, 199 (1996).
\bibitem{iwasaki}
M.~Iwasaki, these proceedings.
\bibitem{macesanu}
C.~Macesanu, these proceedings.
\bibitem{corcella}
G.~Corcella, these proceedings.
\bibitem{sjostrand}
T.~Sjostrand, these proceedings.
\bibitem{wackeroth}
D.~Wackeroth, these proceedings.
\bibitem{real}
C.~Macesanu and L.~H.~Orr, hep-ph/0012177.
\bibitem{boos}
E.~Boos, these proceedings.




\end{references}
\end{document}